\documentclass{aastex62}

\newcommand\about  {\hbox{$\sim$}}

\newcommand\kms    {\hbox{km\,s$^{-1}$}}

\newcommand\mone   {\hbox{$^{-1}$}}

\newcommand\water     {\hbox{H$_2$}O}

\graphicspath{{./}{figures/}}

\accepted{\apjl}

\shorttitle{Molecular Torus of NGC~1068}
\shortauthors{Impellizzeri et al.}

\begin{document}
\title{Counter-Rotation and High Velocity Outflow in the Parsec-Scale Molecular Torus of NGC~1068}

\correspondingauthor{C. M. Violette Impellizzeri}
\email{Violette.Impellizzeri@alma.cl}

\author[0000-0003-4561-1713]{C. M. Violette Impellizzeri}
\affil{Joint ALMA Observatory, Alonso de Cordova 3107, Vitacura, Santiago, Chile}
\affil{National Radio Astronomy Observatory, 520 Edgemont Road, Charlottesville, VA 22902, USA}

\author[0000-0002-6972-2760]{Jack F. Gallimore}
\affil{Department of Physics and Astronomy, 
Bucknell University, 
Lewisburg, PA 17837, USA}

\author[0000-0002-4735-8224]{Stefi A. Baum}
\affil{University of Manitoba, Department of Physics and Astronomy, Winnipeg, MB R3T 2N2, Canada}

\author[0000-0001-8143-3550]{Moshe Elitzur}
\affil{Astronomy Dept., University of California, Berkeley, CA 94720, USA}

\author[0000-0003-4949-7217]{Richard Davies}
\affil{Max Planck Institute for Extraterrestrial Physics, Giessenbachstrasse 1, D-85748 Garching, Germany}

\author[0000-0003-0291-9582]{Dieter Lutz}
\affil{Max Planck Institute for Extraterrestrial Physics, Giessenbachstrasse 1, D-85748 Garching, Germany}

\author[0000-0002-4985-3819]{Roberto Maiolino}
\affil{Kavli Institute for Cosmology, University of Cambridge, Madingley Road, Cambridge CB3 0HA, UK}

\author[0000-0002-9889-4238]{Allesandro Marconi}
\affil{Dipartimento di Fisica e Astronomia, Universit`a di Firenze, via G. Sansone 1, 50019, Sesto Fiorentino (Firenze), Italy}
\affil{INAF-Osservatorio Astrofisico di Arcetri, Largo E. Fermi 5, 50135, Firenze, Italy}

\author[0000-0002-7052-6900]{Robert Nikutta}
\affil{National Optical Astronomy Observatory, 950 N Cherry Ave, Tucson, AZ 85719, USA}

\author[0000-0001-6421-054X]{Christopher P. O'Dea}
\affil{University of Manitoba, Department of Physics and Astronomy, Winnipeg, MB R3T 2N2, Canada}

\author[0000-0002-3140-4070]{Eleonora Sani}
\affil{European Southern Observatory, Alonso de Cordova 3107, Vitacura, Santiago, Chile}

\begin{abstract}
We present 1.4~pc resolution observations of \added{256~GHz nuclear radio continuum and} HCN ($J=3 \to 2$) in the molecular torus of NGC~1068.  \added{The integrated radio continuum emission has a flat spectrum consistent with free-free emission and resolves into an X-shaped structure resembling an edge-brightened bicone.} \added{HCN is detected in absorption against the continuum, and} the \deleted{nuclear} absorption spectrum shows a pronounced blue wing that suggests a high-velocity molecular outflow with speeds reaching 450~\kms. Analysis of the off-nucleus emission line kinematics and morphology reveals two nested, rotating disk components. The inner disk, inside $r\sim 1.2$~pc, has kinematics consistent with the nearly edge-on, geometrically thin \water\ megamaser disk in Keplerian rotation around a central mass of $1.66\times 10^7\,\mbox{M}_\odot$. The outer disk, which extends to $\sim 7$~pc radius, counter-rotates relative to the inner disk. The rotation curve of the outer disk is consistent with rotation around the same central mass as the megamaser disk but in the opposite sense. The morphology of the molecular gas is asymmetric around the nuclear continuum source. We speculate that the outer disk formed from more recently introduced molecular gas falling out of the host galaxy or from a captured dwarf satellite galaxy.  In NGC~1068, we find direct evidence that the molecular torus consists of counter-rotating and misaligned disks on parsec scales.

\end{abstract}

\section{Introduction} \label{sec:intro}
The unifying model for active galaxies holds that all active galactic nuclei (AGNs) are powered by accretion onto a supermassive black hole (SMBH) \citep{1993ARA&A..31..473A,recentReview,2018ARA&A..56..625H}. The accretion disk and broad-line region (BLR) are surrounded by a dusty, molecular medium, commonly referred to as the ``obscuring torus.'' The torus selectively obscures sight-lines to the central engine, resulting in a dichotomy of AGN spectra. If the torus is viewed more nearly pole-on, we can see the BLR and classify the AGN as type~1; if the torus is viewed more nearly edge-on, the BLR may be obscured and the AGN is classified as type~2 \citep{1982ApJ...256..410L}. A parsec-scale torus must be geometrically thick to explain the large detection fraction of type~2 AGNs \citep{1988ApJ...327...89O,2001AN....322...87T}.

NGC~1068 is one of the nearest, luminous Seyfert type 2 galaxies and has served as an archetype for unifying schemes \citep{1985ApJ...297..621A}. Polarimetry of the reflected nuclear spectrum reveals the hidden BLR and UV continuum source \citep{1985ApJ...297..621A,1995MNRAS.275..398I,1995ApJ...452L..87C}.
The Seyfert nucleus of NGC~1068 is obscured by a dusty, molecular medium \citep{2006ApJ...640..612M,2016ApJ...823L..12G,2016ApJ...829L...7G,2018ApJ...853L..25I}.  A flat spectrum radio source, ``S1,'' marks the location of the central engine \citep{1996MNRAS.278..854M,1996ApJ...458..136G}. VLBA continuum observations resolve the continuum source into an elongated, pc-scale structure oriented nearly at right angles to the kpc-scale radio jet; the radio source has been interpreted as a plasma disk located inside the obscuring molecular medium \citep{1997Natur.388..852G,2004ApJ...613..794G}. The cm-wave continuum source associates with the disk of H$_2$O vapor megamaser emission with systemic velocity 1130~\kms\ (optical convention, LSRK reference frame) and maximum rotation speed $335$~\kms\ at a distance of 0.6~pc from the central engine \citep{1997Ap&SS.248..261G,2001ApJ...556..694G,newmaserpaper}. 

Recent ALMA observations of the molecular interstellar medium (ISM) in the central few pc have revealed complicated kinematics \citep{ 2016ApJ...823L..12G,2016ApJ...829L...7G,2018ApJ...853L..25I}. With a $\sim$~3~pc (0.04\arcsec) beam\footnote{We adopt the Ringberg Standards for NGC~1068 \citep{1997Ap&SS.248....9B}, which assigns the scale 1\arcsec = 70~pc.}, ALMA observations of CO ($J=6\to 5$) and HCN ($J=3\to 2$) resolve an extended, roughly 10~pc by 4~pc distribution of molecular gas oriented nearly at right angles to the radio jet axis but nearly aligned with the major axis of the nuclear plasma disk, S1. We found evidence for high velocity CO in a bipolar outflow, with radial velocities ranging to $\pm 400$~\kms\ relative to systemic, in the central pc \citep{2016ApJ...829L...7G}. Imanishi et al. observed HCN ($J=3\to 2$) at $\sim 3$~pc resolution, and they argued that, unlike CO ($J=6\to 5$), the molecular gas traced by HCN better traces the rotational, disk-like region of the torus. However, they also argue that the kinematics are not consistent with pure rotation but show evidence for significant turbulence (i.e., enhanced velocity dispersion revealed by the second spectral moment) and perhaps counter-rotation compared to the molecular gas on hundred parsec scales.

In an effort to untangle the complicated geometry and kinematics of the molecular torus of NGC~1068, we obtained new ALMA observations of HCN ($J=3\to 2$) at 1.4~pc (20~mas) resolution. The observations and data reduction are summarized in Section~\ref{sec:obs}. The main results, including moment maps and a kinematic analysis, are presented in Section~\ref{sec:results}. We summarize our main conclusions in Section~\ref{sec:conclusions}.

\section{Observations and Data Reduction} \label{sec:obs}

We used ALMA to observe NGC~1068 in Band~6 during ALMA Cycle~5 (project code 2017.1.01666.S). We used the longest baseline configuration available for Band~6 during Cycle~5, C43-10, with baselines ranging from 41.4~m to 16.2~km. The largest recoverable structure is about 0\farcs32, or 22~pc at the distance of NGC~1068. The source was observed in nine execution blocks between 10 October 2017 and 24 October 2017. The average exposure time per execution block was 1.6~hours, and the total integration time on source was 14.5~hours. Source observations were interleaved with short scans of nearby bright
calibrator sources. The phase calibrator was J0239$-$0234, located 2\fdg7 from NGC~1068, and the flux calibrator was J0432$-$0120, with flux density $S_{\nu} = 0.97\pm0.04$~Jy.

Spectral windows were tuned to observe three continuum windows, 1.875~GHz total bandwidth each and one tuned to observe HCN ($J = 3\to 2$; $\nu_0 = 265.88618$~GHz) at recessional velocity 1150~\kms\ (optical convention, LSRK reference frame). The HCN line observation has channel-width \about 1.0~\kms\ and spans \about 1000~\kms\ total bandwidth. 

The data were calibrated using CASA pipeline version 40896 (Pipeline-CASA51-P2-B) and imaged using CASA 5.1.1 \citep{2007ASPC..376..127M}.  We performed four cycles of phase-only self-calibration on the continuum data \citep{1989ASPC....6..185C}, and the solutions were applied to the spectral line data.  The visibility data were Briggs-weighted during Fourier inversion \citep{BriggsWeighting}. The spectral line data were averaged in frequency to provide approximately 10~\kms\,wide channels. Continuum subtraction of the line data was performed in the $(u,v)$ plane using an average of channels with radial velocities $| \mbox{RV}| > 600$\,\kms{} relative to systemic. The common restoring beam is $19.9\times19.1$~mas, PA~79\fdg8, and the median line channel sensitivity is 1.6~mJy~beam$^{-1}$. 

The continuum image was produced by combining all continuum channels into a single image at effective $\nu=256.5$~GHz (multifrequency synthesis; \citealt{1999ASPC..180..419S}). The restoring beam of the continuum image is $20.2\times19.6$~mas, PA~78\fdg3, and the rms sensitivity is $6.2\,\mu\mbox{Jy\ beam}^{-1}$. 

\section{Results}\label{sec:results}

\begin{figure*}[tbh]
\centering
\plotone{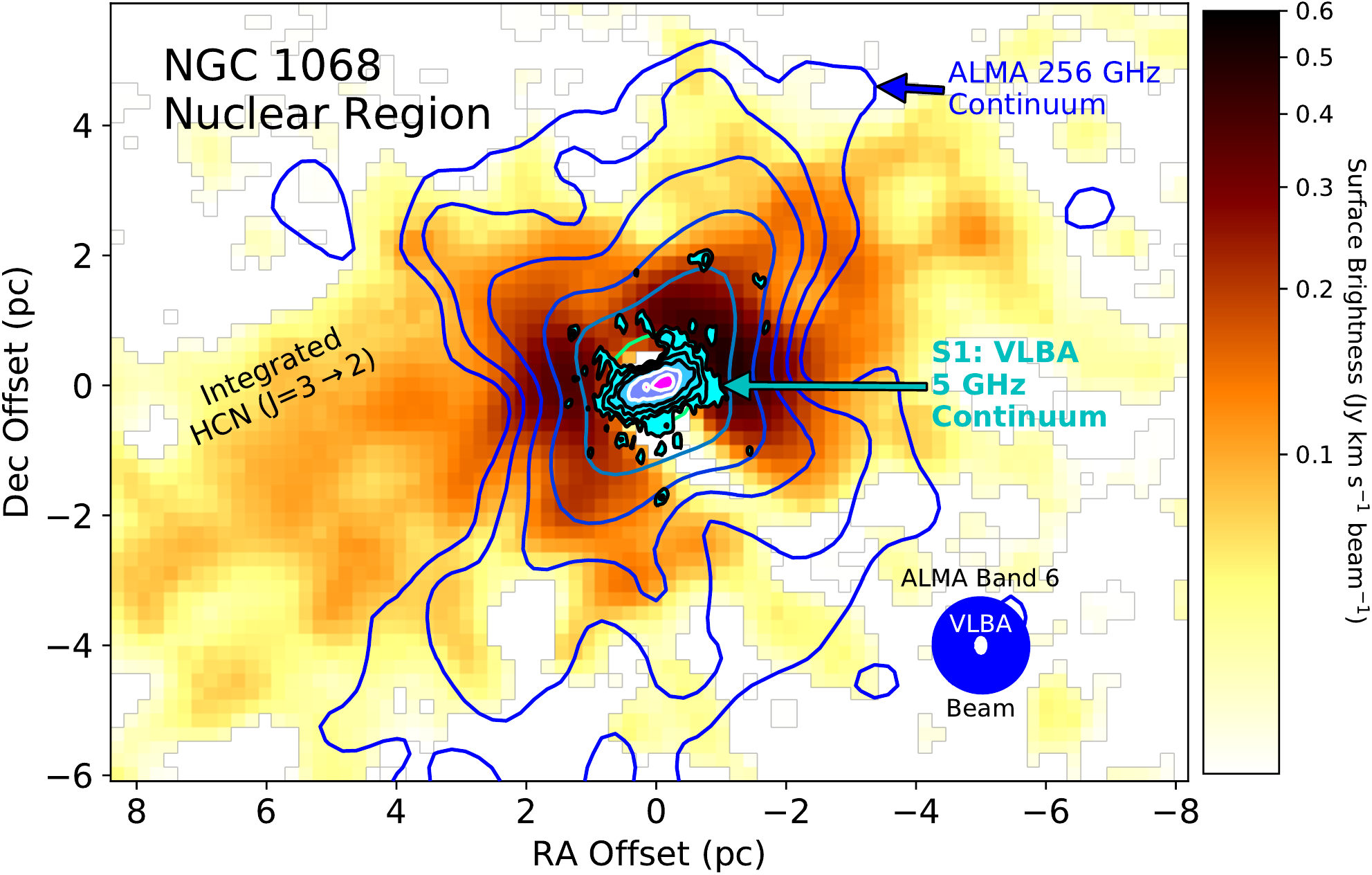}
\caption{\explain{This figure was modified to include a colorbar to the right of the main plot.} The nuclear region of NGC~1068 shown as overlays of the ALMA 256~GHz continuum (blue-green unfilled contours; Band~6), VLBA 5~GHz continuum (black contours filled with shades of cyan to magenta; \citealt{2004ApJ...613..794G}), and ALMA integrated HCN ($J=3\rightarrow2$; Band~6) (background image in shades of yellow to dark red). A compact source model for S1 has been subtracted from the ALMA continuum; the flux density of the compact source is $6.6\pm 0.3$~mJy. The Band~6 contour continuum levels are (blue) 0.019, 0.037, 0.075, 0.15, (blue-green) 0.30, and (green, mostly obscured by the VLBA continuum contours) 0.60~mJy~beam\mone. \replaced{The integrated HCN image is shown with an arcsinh stretch and surface brightness limits 0 to 0.6~Jy~\kms~beam\mone.}{The integrated HCN image is shown with an arcsinh stretch and surface brightness limits 0.004 ($5\sigma$) to 0.6~Jy\,\kms\,beam\mone. Blanked values ($< 5\sigma$) display as white.} }\label{fig:integrated}
\end{figure*}

Figure~\ref{fig:integrated} presents images of the HCN ($J=3 \to 2$) integrated line flux and 256~GHz continuum. The line is detected in absorption at the position of S1, and the brightest integrated emission extends about 7~pc from S1 both southeast and northwest of S1. ALMA resolves the integrated HCN emission, with surface brightness (uncorrected for absorption) peaking roughly 2~pc to either side of S1 and fading with distance from S1. The morphology is irregular, suggesting substructure on scales at least as small as the $\sim 1.4$~pc beam. The orientation of the HCN major axis is PA~$\sim 114\degr$, estimated by drawing a line through the continuum peak and between the outer contours of the integrated HCN flux map. \added{Image moment analysis agrees with this estimate of the major axis orientation to within \about 5\degr, which corresponds to a difference of less than half of a beam at the extent of the major axis.} The major axis is more closely aligned with the milliarcsecond-scale radio continuum structure of S1 but nearly perpendicular to the radio jet on 100~pc scales.

\begin{figure*}[tbh]
\centering
\plotone{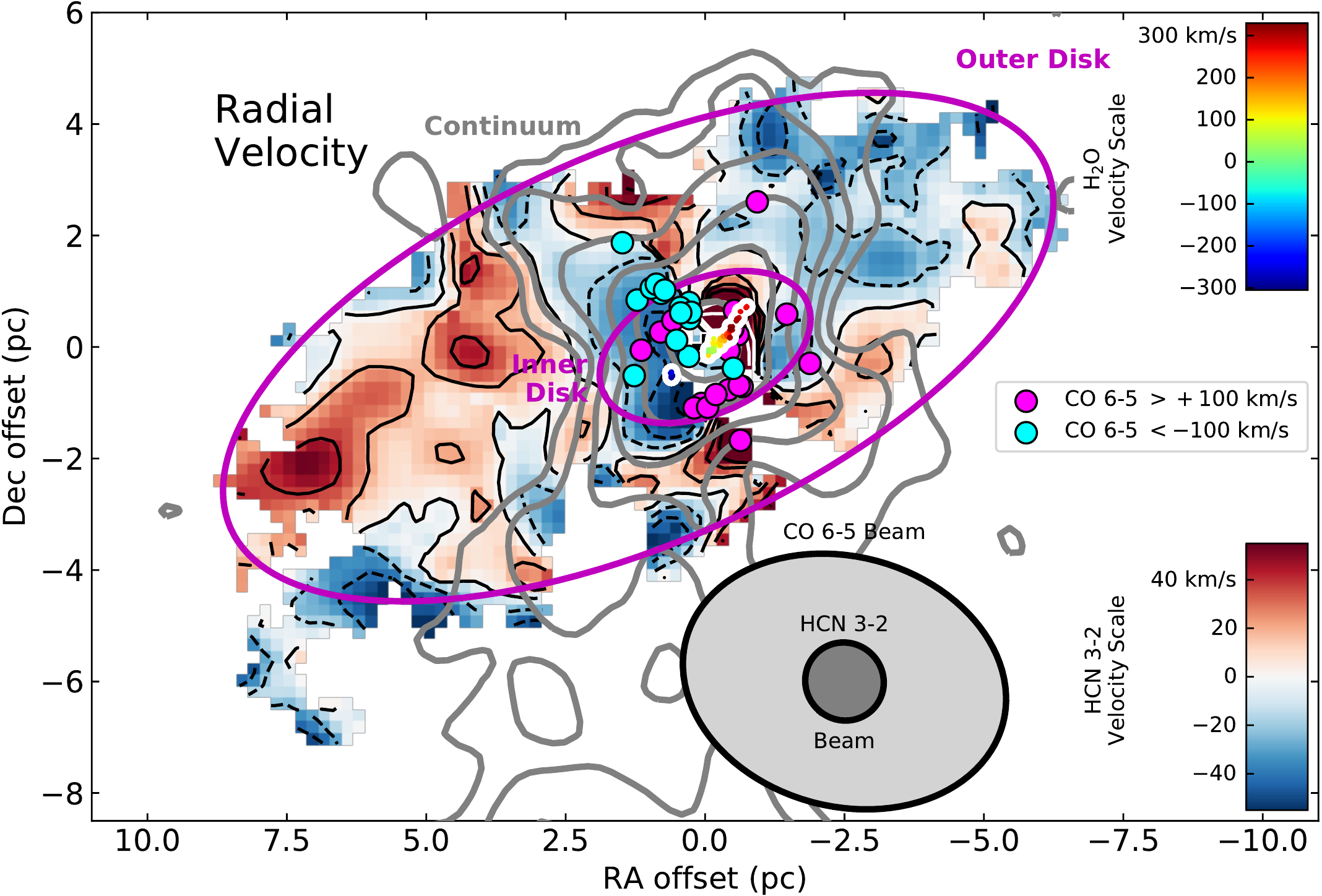}
\caption{\explain{We replaced the elliptical outflow annotation with radio continuum contours. In hindsight, although it makes the figure busy, the continuum contours better represent a projection of the putative outflow region. We also rescaled the HCN 3-2 velocity scale slightly to bring out the contrast of the east and west sides of the velocity map.} Radial velocities of molecular gas in the nuclear region of NGC~1068. The systemic velocity, 1130~\kms\ (LSRK, optical convention; \citealt{newmaserpaper}), has been subtracted from the radial velocities. Included in this figure are the H$_2$O megamaser spots (colored spots outlined in white; \citealt{1997Ap&SS.248..261G}), \added{the 256~GHz continuum (gray contours as in Figure~\ref{fig:integrated})}, the velocity map of HCN $J=3 \rightarrow 2$ (black contours with red--blue shading), and spots marking the peak surface brightness emission of high-velocity CO ($J=6\rightarrow 5$) \citep{2016ApJ...829L...7G} (magenta and cyan filled circles). The velocity contours run from $-60$~\kms\ to $+60$~\kms\ in steps of 20~\kms; negative velocities are contoured with dashed lines. The regions identified as the inner and outer counter-rotating disks are outlined by annotated, magenta ellipses. \deleted{The outflow region is similarly outlined and annotated in cyan.} }\label{fig:velocity}
\end{figure*}

At lower surface brightness levels, the ALMA 256~GHz continuum image resolves into an X-shaped structure that resembles a bicone extending perpendicular both to the molecular gas distribution and the major axis of the VLBA continuum source. We have interpreted the VLBA continuum source as arising from a parsec-scale plasma disk \citep{2004ApJ...613..794G}. That the major axis of the \replaced{ALMA continuum image}{fainter ALMA continuum emission} aligns nearly perpendicular to the VLBA major axis is consistent with a disk-bicone scenario. On the sub-pc scale, the VLBA continuum contours also show X-shaped extensions away from the plasma disk plane that appear to line up roughly with the ALMA continuum extensions. \added{Although a detailed analysis of the continuum is outside the scope of this {\em Letter}, we note that the 256~GHz continuum is probably dominated by free-free emission. The integrated flux density of the S1 region in a 0\farcs13 (9.1~pc) square aperture is $12.7\pm 0.1$~mJy. Based on the VLA~43~GHz measurement of \cite{2008A&A...477..517C} and assuming a 5\% uncertainty in the flux scale, the continuum spectral index is $\alpha = -0.06\pm0.03$ ($S_{\nu} \propto \nu^{\alpha}$; $1\sigma$ uncertainty). Accepting additional uncertainty owing to the difference in resolution (80~mas for the VLA) and flux scale calibration, this measurement is consistent with the spectral index expected for optically-thin free-free emission, $\alpha=-0.1$.}

We measured the radial velocity field \added{of the spectral line cube} using Gauss-Hermite polynomial fits to each spectrum in the HCN data cube (see \citealt{1993ApJ...407..525V} for a description of this technique); the result is shown in Figure~\ref{fig:velocity}. Outside of about 1~pc projected radius, the broad trend is redshifts relative to systemic on the eastern side and blueshifts on the western side. The kinematics are consistent with the results presented by \cite{2018ApJ...853L..25I} but opposite the sense of motion of the H$_2$O masers \citep{1997Ap&SS.248..261G}. Inside the central parsec, however, the kinematics more closely resemble the \water\ megamaser disk, with high velocity redshifted HCN northwest of the nucleus and high velocity blueshifted HCN southeast of S1. 

The rotating disk(s) interpretation is complicated by the velocity field northeast and southwest of the nucleus, which does not cleanly resolve from the near-nuclear gas. This axis coincides with the bipolar outflow seen in CO ($J=6\rightarrow 5$;  \citealt{2016ApJ...829L...7G}), and so we speculate the complex radial velocity field northeast and southwest of the nucleus results from a beam-convolved blend of clouds in the rotating disk and clouds participating in the outflow. Supporting this picture, we find that the high-velocity HCN ($J=3 \to 2$) appears to bracket the CO ($J=6\rightarrow 5$) outflow. The simplest interpretation would have the HCN trace molecular gas in a rotating disk that collimates the molecular outflow detected in CO ($J=6\rightarrow 5$) emission and evidenced by the disturbed HCN radial velocities along the minor axis. 

\begin{figure*}[tbh]
\centering 
\plotone{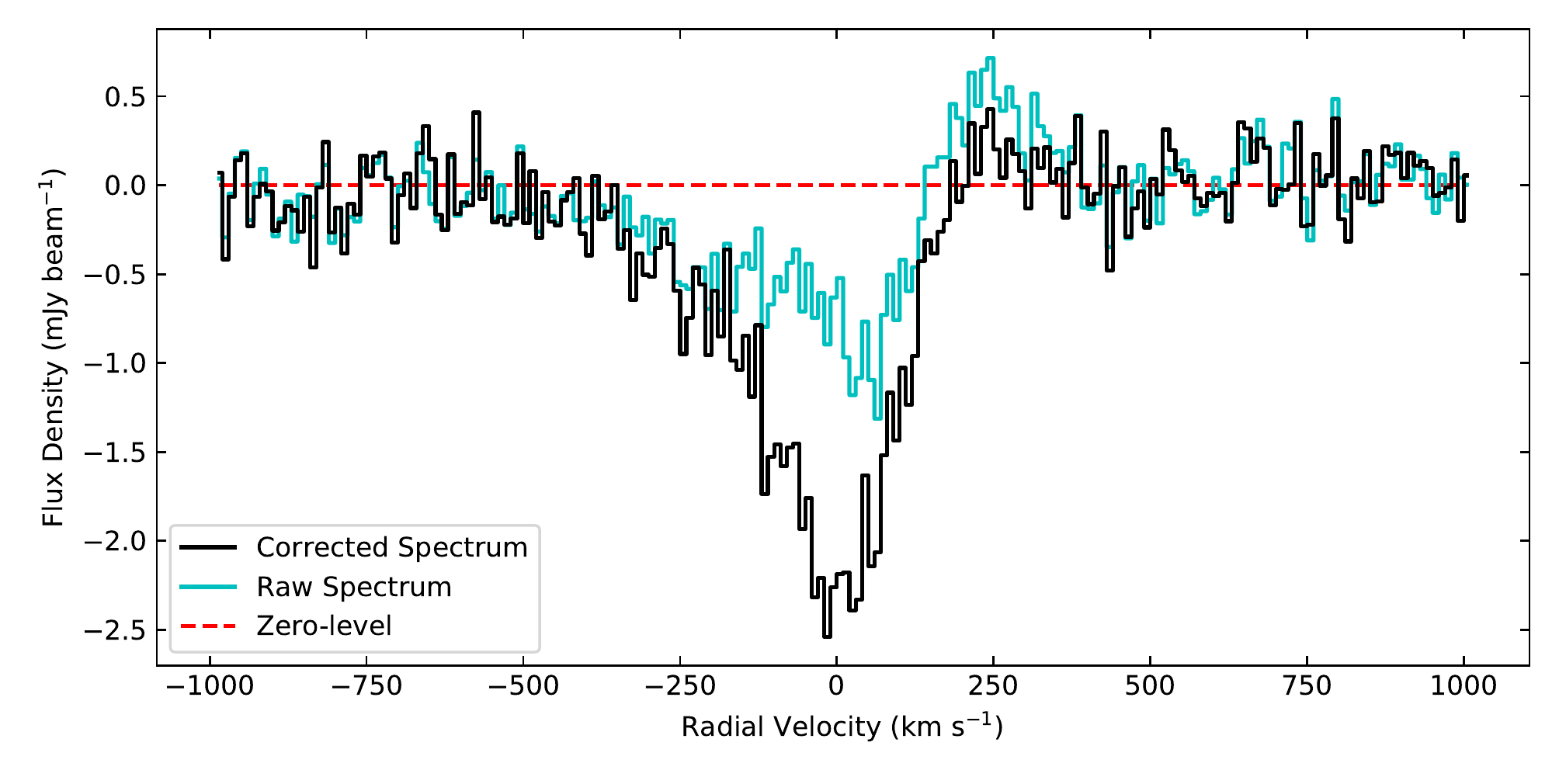}
\caption{The continuum-subtracted nuclear absorption profile of HCN ($J=3\rightarrow 2$). The raw spectrum is plotted in cyan, the spectrum corrected for line emission is plotted in black, and the red dashed line shows the zero-level. The systemic velocity, 1130~\kms\ (LSRK, optical convention; \citealt{newmaserpaper}), has been subtracted from the radial velocities.}
\label{fig:absorption}
\end{figure*}

The nuclear absorption profile, provided in  Figure~\ref{fig:absorption}, shows additional evidence for an outflow component in HCN ($J=3 \to 2$). We extracted the nuclear spectrum at the position of the nuclear radio source S1. To estimate a correction for line emission in the beam, we averaged the spectra from positions 1.4~pc (one beam) northeast and southwest of the nucleus and subtracted the result from the nuclear spectrum. The resulting, corrected absorption profile shows a minimum centered near the systemic velocity, as expected for an edge-on, rotating disk surrounding the nuclear continuum source. However, the line profile is asymmetric: there is a pronounced blue wing on the absorption profile that extends to roughly $-450$~\kms\ relative to systemic. The blue wing appears on both the raw and corrected spectra, and so it is likely not an artifact of the emission correction. For comparison, the maximum rotation speed of the inner disk is $\sim 380$~\kms (see below). The excess blueshifted absorption is consistent with outflow velocities approaching 450~\kms, exceeding the 400~\kms, maximum outflow velocities observed in CO ($J=6\rightarrow 5$) line emission \citep{2016ApJ...829L...7G}. 

\begin{figure*}
\centering 
\plotone{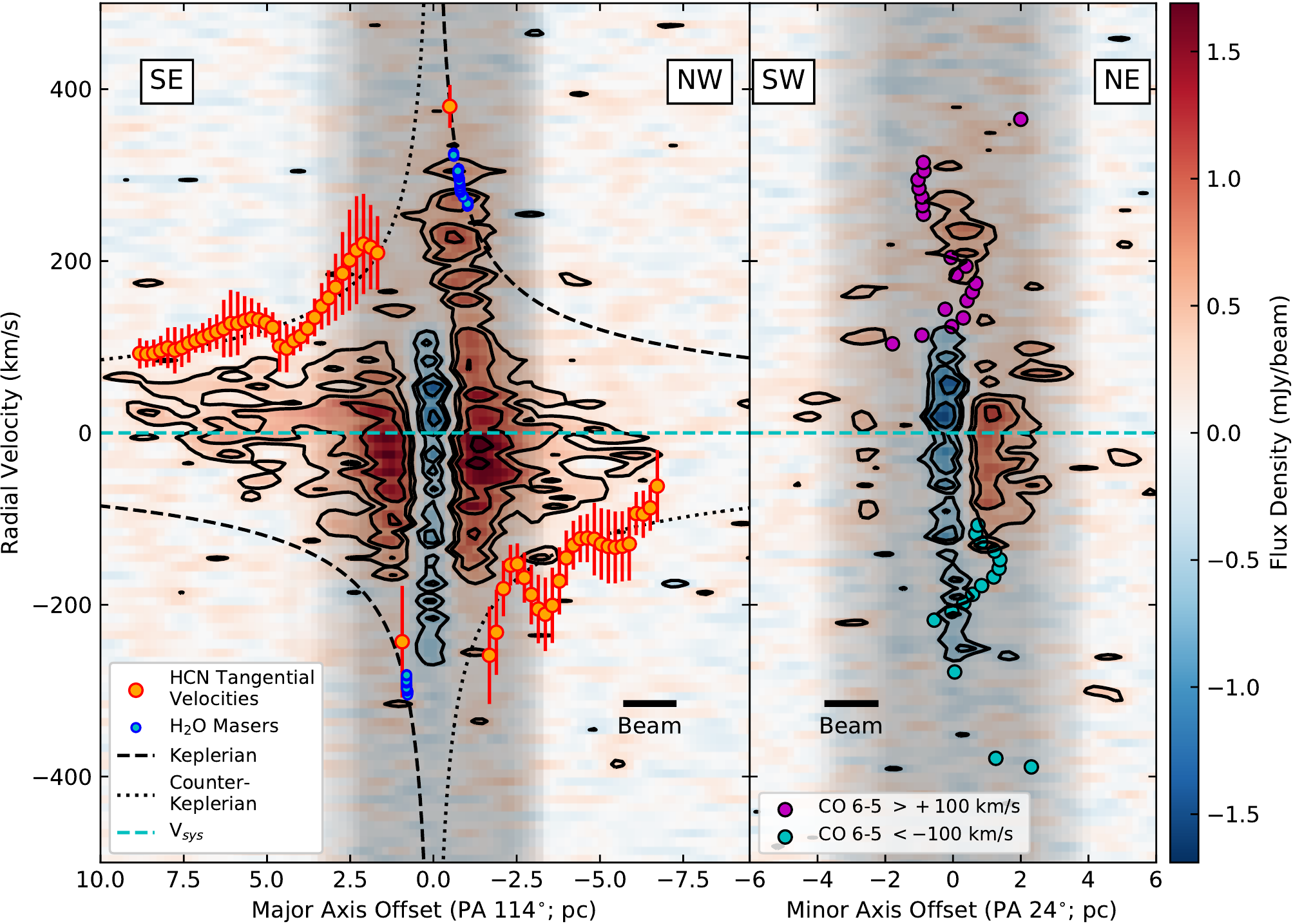}
\caption{\explain{This figure was modified in several ways based on the referee's suggestions. The tangential velocity curves are now superimposed on position-velocity diagrams and moved to the left panel. The new, right panel has been added to show the minor axis position-velocity diagram.} \replaced{ALMA HCN ($J=3\to2$) tangential velocity curve of the central \about 8~pc of NGC~1068.}{ALMA HCN ALMA HCN ($J=3\to2$) position-velocity diagrams extracted along the HCN major axis (left panel) and minor axis (right panel).} The radial velocities are offset relative to systemic, 1130~\kms (LSRK, optical convention; \citealt{newmaserpaper})\added{, which is marked by the blue-green dashed line}. \replaced{The major axis offsets run along PA~$114\degr$ through the nuclear continuum source S1. Positions run southeast (positive offset) through northwest (negative offset).}{The major axis offsets run southeast (positive offset) to northwest (negative offset) along PA~$114\degr$ through the nuclear continuum source S1, and the minor axis offsets run southwest (negative offset) to northeast (positive offset) along PA~$24\degr$.} \added{The contour levels run from $-1.6$ to 1.6~mJy\,beam\mone\ in steps of 0.4~mJy\,beam\mone.}
\added{On the major axis diagram,} the tangential HCN velocities are marked by yellow dots with red errorbars, and, for comparison, the H$_2$O maser radial velocities along PA \replaced{$\sim 135\degr$}{$\sim 131\degr$} \explain{Woops, typo.} are shown as filled blue circles (data from \citealt{1997Ap&SS.248..261G}) corrected for the best-fit inclination, $i = 79\degr$. The best-fitting model for the H$_2$O masers, shown with a dashed line, is Keplerian rotation around a central mass of $1.66\times 10^7 \mbox{M}_\odot$ (inner disk rotation; \citealt{newmaserpaper}).  The dotted line shows the same Keplerian rotation, but in the opposite sense. \added{The 256~GHz continuum emission is indicated by the shaded gray regions. The grayscale stretch runs between 16 to 50~mJy~beam\mone. }}
\label{fig:tanvel}
\end{figure*}

\explain{Our goal with this paper was to emphasize the rotation of the outer disk rather than the outflow. As requested by the referee, we have now included position-velocity diagrams that contain more information about the outflow region. We have added the following paragraphs to discuss the position velocity diagrams.}
\added{Position-velocity diagrams along the major (PA~114\degr) and minor (PA~24\degr) axes are shown in Figure~\ref{fig:tanvel}. On the major axis diagram, the highest velocity emission agrees with the positional offsets and range of velocities spanned by the redshifted H$_2$O masers (see also Figure~\ref{fig:velocity}) and so likely arises from the inner disk. Within offsets $\la 3$~pc, the region of the brightest continuum emission, the brightest HCN line emission spans radial velocities $-170$ to $100$~\kms relative to systemic. Given the coincidence with the radio continuum and the presence of both redshifted and blueshifted emission, we interpret this region as including a significant contribution from spatially unresolved molecular outflow, consistent with our interpretation of the velocity map (Figure~\ref{fig:velocity}). The outflow emission is most likely overlapping with some of the higher velocity rotating gas. The presence of both redshifted and blueshifted emission suggests a biconical outflow with an outflow axis aligned near the plane of the sky (cf. \citealt{2006AJ....132..620D}).
For offsets $\ga 4$~pc, the emission is preferentially redshifted to the southeast and blueshifted to the northwest, as is evident on the velocity map (Figure~\ref{fig:velocity}). This region appears to be well-separated from the outflow region and runs roughly orthogonal to the known outflow and jet axes. Although some beam-smeared outflow emission may contribute to this outer region, it seems more likely that rotation dominates the kinematics of the outer part of the HCN emission along the major axis.

The minor axis position-velocity diagram places some constraints on the scale of the outflow. Allowing for beam-smearing, HCN ($J = 3\to 2$) emission extends roughly 1~pc northeast of the nucleus, but the HCN  emission is much weaker to the southwest (cf. Figure~\ref{fig:integrated}). The CO ($J = 6\to 5$) emission similarly spans roughly 1--2~pc both southwest and northeast of the nucleus. Of course, the presence of line absorption, sensitivity limitations, and, potentially, confusion with overlapping emission from the inner disk complicate the measurement of the extent of the molecular outflow in HCN. On the other hand, molecular gas lifted from the shielding of the nuclear disk(s) would be soon exposed to the ionizing continuum from the AGN, and we speculate that the molecular outflow might become ionized and feed into to the larger, $\sim 400$~pc ionized gas outflow (cf. \citealt{1990ApJ...355...70C}). The implication would be acceleration from maximum LOS velocities of roughly 450~\kms\ for outflowing molecular clouds (Figure~\ref{fig:absorption}) to 1000~\kms\ for [OIII] emission line clouds within the central 14~pc \citep{2006AJ....132..620D}.} 

We measured the parsec-scale rotation curve from the HCN ($J = 3\to 2$) data using the conventional tangential velocity technique \citep{1965gast.book..167K}. To perform this analysis, we used an automated, numerical technique to determine the tangential velocity as a function of offset along the major axis of the integrated HCN map. For projected $r \ge 1.2$~pc, we extracted spectra at each pixel along the major axis (1~pixel = 0.2~pc). For $r < 1.2$~pc, we instead extracted two spectra, one northwest of the nucleus and one southeast along the axis defined by the maser spot distribution (PA~131\degr) and at the locations of the brightest peak line emission in that region. The northeast spectrum was extracted at a position located $r = 0.5$~pc from the nuclear continuum source. Owing to confusion with the blue wing of the nuclear absorption line, the extraction for the southeast position is farther from the nucleus, at projected radius $r = 0.9$~pc.

A Gauss-Hermite polynomial was fitted to each extracted HCN spectrum, and Monte Carlo realizations of the spectrum were generated by adding a randomly generated noise spectrum scaled to the observed rms. The procedure next bracketed the region surrounding the spectral peak and the nearest velocity where the simulated spectrum falls below the background noise. The average and standard deviation of $10^4$ trials respectively provide an estimate of the tangential velocity and its uncertainty. The median uncertainty is 30~\kms.  Note that this method does not correct for the unknown, intrinsic velocity dispersion of the gas. However, intrinsic velocity dispersions as high as $\sigma_r = 90$~\kms, comparable to the inferred rotation speeds, introduce corrections smaller than the measurement uncertainties.

The results are shown in Figure~\ref{fig:tanvel}, which also includes the radial velocities of the nuclear \water\ megamasers and the Keplerian rotation curve that best fits the \water\ megamaser kinematics \citep{newmaserpaper}.

The $r > 1.2$~pc rotation curve is consistent with an extrapolation of the Keplerian rotation curve. There are bumps and wiggles that deviate slightly from Keplerian rotation \added{perhaps owing to confusion with the molecular outflow, i.e. overlap between the outflow and rotating gas within our beam}; however, from inspection of Figure~\ref{fig:tanvel}, these deviations are small compared to the measurement uncertainties. More specifically, the largest statistical deviation from the predicted Keplerian rotation curve is located 3.4~pc northwest of the nucleus. There, the measured rotational speed deviates from the predicted Keplerian rotation curve by $70\pm 40$~\kms, which is a $< 2\sigma$ deviation.

Again we emphasize that, between $r = 1.2$~pc and $r \sim 7$~pc, the outer HCN rotation curve runs counter to the inner maser disk rotation; that is, the masers recede toward the northwest, but the outer HCN disk recedes to the southeast. However, inside $r = 1.2$~pc, the HCN rotation follows the sense of the maser disk rotation. The inferred rotation speeds of the inner HCN disk reach $380\pm 30$~\kms, exceeding the maximum rotation speed of the \water\ megamaser disk, 335~\kms. This result suggests that we have detected molecular gas closer to the central engine than the gas traced by the  \water\ megamasers. Note also that the masers align along PA~131\degr\ (from the nucleus toward the approaching masers; \citealt{newmaserpaper}), misaligned from the integrated, larger-scale HCN major axis by nearly 20\degr.

\section{Conclusions}\label{sec:conclusions}

We used ALMA to observe \added{256~GHz continuum and} the transition HCN ($J=3\to 2$) at roughly parsec resolution in the molecular torus of NGC~1068.  \added{Comparing to VLA 43~GHz measurements, the continuum has a flat spectrum consistent with free-free emission. The continuum structure resembles a bicone viewed from the side, perpendicular to the symmetry axis.}

\added{Turning to the HCN observations,} we identify three kinematically distinct regions: (1) an outflow component \added{apparent in emission on the HCN position-velocity diagram and} detected \added{as a blueshifted wing} in absorption against the nuclear continuum source, with projected outflow speeds approaching $\sim 450$~\kms; (2) an inner disk spanning $0.5 \la r \la 1.2$~pc; and (3) an outer disk extending to $r \sim 7$~pc. The two disks counter-rotate, and the kinematics of the inner disk agree with the H$_2$O megamaser disk mapped by the VLBA \citep{1997Ap&SS.248..261G}. The outer disk shows a Keplerian rotation curve consistent with an extrapolation of the rotation curve of the inner disk. We also find that the HCN radial velocity field is more complex along the molecular outflow axis, which suggests that we are also detecting but not fully resolving HCN emission associated with the outflow.

Conservation of angular momentum precludes the formation of counter-rotating disks out of a single epoch of accreting gas, and, to our knowledge, accretion disk warping mechanisms cannot induce counter-rotation (e.g., \citealt{2014MNRAS.441.1408T} and references therein). It seems likely that the counter-rotating outer disk formed from gas that has been more recently introduced to the $ > 1$~pc-scale environment, perhaps from an errant cloud falling out of the host galaxy or a captured satellite dwarf galaxy. \cite{2018ApJ...853L..25I} demonstrated that the outer disk of the parsec-scale molecular torus also counter-rotates with respect to the molecular ring at 100~pc scales (see their Figure~1) and, we note, the sense of rotation of the galactic disk on kiloparsec scales (see, e.g., \citealt{1995ApJ...450...90H} and \citealt{1997Ap&SS.248...23B}).

\added{Since the central, supermassive black hole appears to dominate the gravitational potential in this region, the counter-rotating disks should be stable until, owing to dissipation, the outer disk starts to collapse onto the inner disk. When that occurs, we would expect the outer disk to collapse on an orbital timescale, $t_{orb} \sim 3\times10^4$~years.}
The interaction between counter-rotating disks may enhance the accretion rate by one or two orders of magnitude \citep{1999ApJ...514..691K,2012MNRAS.422.2547N,2015MNRAS.446..613D}.  \cite{2006MNRAS.373L..90K,2007MNRAS.377L..25K} describe a chaotic accretion scenario for the cosmological growth of black holes, namely, rapid accretion owing to randomly oriented and episodic accretion events. They cite the random orientation of Seyfert radio jets and ionization cones with respect to host galaxies as evidence for chaotic accretion (cf. \citealt{2003ApJ...597..768S, 2006AJ....132..546G}). In NGC~1068, we find direct evidence that the molecular obscuring medium consists of counter-rotating and misaligned disks on parsec scales.

\acknowledgments

We thank \added{Santiago Garc{\'\i}a-Burillo, the referee,} and the participants of the TORUS 2018 workshop (Puerto Varas, Chile; December 2018), where we first presented the discovery of the counter-rotating outer disk, for helpful feedback that improved this work. This paper makes use of the following ALMA data: ADS/JAO.ALMA\#2017.1.01666.S. ALMA is a partnership of ESO (representing its member states), NSF (USA) and NINS (Japan), together with NRC (Canada), MOST and ASIAA (Taiwan), and KASI (Republic of Korea), in cooperation with the Republic of Chile. The Joint ALMA Observatory is operated by ESO, AUI/NRAO and NAOJ. The National Radio Astronomy Observatory is a facility of the National Science Foundation operated under cooperative agreement by Associated Universities, Inc.

\facility{ALMA}

\software{CASA \citep{2007ASPC..376..127M}, astropy \citep{2013A&A...558A..33A}}





\begin{thebibliography}{}

\bibitem[Antonucci \& Miller(1985)]{1985ApJ...297..621A} Antonucci, R.~R.~J., \& Miller, J.~S.\ 1985, \apj, 297, 621 

\bibitem[Antonucci(1993)]{1993ARA&A..31..473A} Antonucci, R.\ 1993, \araa, 31, 473 


\bibitem[Astropy Collaboration et al.(2013)]{2013A&A...558A..33A} Astropy Collaboration, Robitaille, T.~P., Tollerud, E.~J., et al.\ 2013, \aap, 558, A33 

\bibitem[Bland-Hawthorn et al.(1997)]{1997Ap&SS.248....9B} Bland-Hawthorn, J., Gallimore, J.~F., Tacconi, L.~J., et al.\ 1997, \apss, 248, 9 


\bibitem[Briggs(1995)]{BriggsWeighting} Briggs, D. 1995, PhD Thesis, New Mexico Institute of Mining
and Technology

\bibitem[Brinks et al.(1997)]{1997Ap&SS.248...23B} Brinks, E., Skillman, E.~D., Terlevich, R.~J., et al.\ 1997, \apss, 248, 23


\bibitem[Capetti et al.(1995)]{1995ApJ...452L..87C} Capetti, A., Macchetto, F., Axon, D.~J., Sparks, W.~B., \& Boksenberg, A.\ 1995, \apjl, 452, L87 


\bibitem[Cecil et al.(1990)]{1990ApJ...355...70C} Cecil, G., Bland, J., \& Tully, R.~B.\ 1990, \apj, 355, 70

\bibitem[Cornwell \& Fomalont(1989)]{1989ASPC....6..185C} Cornwell, T., \& Fomalont, E.~B.\ 1989, Synthesis Imaging in Radio Astronomy, 6, 185 

\bibitem[Cotton et al.(2008)]{2008A&A...477..517C} Cotton, W.~D., Jaffe, W., Perrin, G., et al.\ 2008, \aap, 477, 517



\bibitem[Das et al.(2006)]{2006AJ....132..620D} Das, V., Crenshaw, D.~M., Kraemer, S.~B., et al.\ 2006, \aj, 132, 620

\bibitem[Dyda et al.(2015)]{2015MNRAS.446..613D} Dyda, S., Lovelace, R.~V.~E., Ustyugova, G.~V., Romanova, M.~M., \& Koldoba, A.~V.\ 2015, \mnras, 446, 613 


\bibitem[Gallimore et al.(1996)]{1996ApJ...458..136G} Gallimore, J.~F., Baum, S.~A., O'Dea, C.~P., \& Pedlar, A.\ 1996, \apj, 458, 136 

\bibitem[Gallimore et al.(1997)]{1997Natur.388..852G} Gallimore, J.~F., Baum, S.~A., \& O'Dea, C.~P.\ 1997, \nat, 388, 852 

\bibitem[Gallimore et al.(2001)]{2001ApJ...556..694G} Gallimore, J.~F., Henkel, C., Baum, S.~A., et al.\ 2001, \apj, 556, 694 

\bibitem[Gallimore et al.(2004)]{2004ApJ...613..794G} Gallimore, J.~F., Baum, S.~A., \& O'Dea, C.~P.\ 2004, \apj, 613, 794 

\bibitem[Gallimore et al.(2006)]{2006AJ....132..546G} Gallimore, J.~F., Axon, D.~J., O'Dea, C.~P., Baum, S.~A., \& Pedlar, A.\ 2006, \aj, 132, 546 

\bibitem[Gallimore et al.(2016)]{2016ApJ...829L...7G} Gallimore, J.~F., Elitzur, M., Maiolino, R., et al.\ 2016, \apjl, 829, L7

\bibitem[Gallimore \& Impellizzeri(2019)]{newmaserpaper} Gallimore, J.~F., \& Impellizzeri, C.~M.~V.\ 2019, \aj, submitted.

\bibitem[Garc{\'{\i}}a-Burillo et al.(2016)]{2016ApJ...823L..12G} Garc{\'{\i}}a-Burillo, S., Combes, F., Ramos Almeida, C., et al.\ 2016, \apjl, 823, L12 
\bibitem[Greenhill \& Gwinn(1997)]{1997Ap&SS.248..261G} Greenhill, L.~J., \& Gwinn, C.~R.\ 1997, \apss, 248, 261 

\bibitem[Helfer \& Blitz(1995)]{1995ApJ...450...90H} Helfer, T.~T., \& Blitz, L.\ 1995, \apj, 450, 90

\bibitem[Hickox \& Alexander(2018)]{2018ARA&A..56..625H} Hickox, R.~C., \& Alexander, D.~M.\ 2018, \araa, 56, 625 

\bibitem[Imanishi et al.(2018)]{2018ApJ...853L..25I} Imanishi, M., Nakanishi, K., Izumi, T., \& Wada, K.\ 2018, \apjl, 853, L25 

\bibitem[Inglis et al.(1995)]{1995MNRAS.275..398I} Inglis, M.~D., Young, S., Hough, J.~H., et al.\ 1995, \mnras, 275, 398 

\bibitem[Kerr \& Westerhout(1965)]{1965gast.book..167K} Kerr, F.~J. \& Westerhout, G.\ 1965, Galactic Structure. Edited by Adriaan Blaauw and Maarten Schmidt. Published by the University of Chicago Press, 167.

\bibitem[King \& Pringle(2006)]{2006MNRAS.373L..90K} King, A.~R., \& Pringle, J.~E.\ 2006, \mnras, 373, L90

\bibitem[King \& Pringle(2007)]{2007MNRAS.377L..25K} King, A.~R., \& Pringle, J.~E.\ 2007, \mnras, 377, L25 

\bibitem[Kuznetsov et al.(1999)]{1999ApJ...514..691K} Kuznetsov, O.~A., Lovelace, R.~V.~E., Romanova, M.~M., \& Chechetkin, V.~M.\ 1999, \apj, 514, 691 

\bibitem[Lawrence \& Elvis(1982)]{1982ApJ...256..410L} Lawrence, A., \& Elvis, M.\ 1982, \apj, 256, 410

\bibitem[Mason et al.(2006)]{2006ApJ...640..612M} Mason, R.~E., Geballe, T.~R., Packham, C., et al.\ 2006, \apj, 640, 612 

\bibitem[{{McMullin} {et~al.}(2007){McMullin}, {Waters}, {Schiebel}, {Young},
  \& {Golap}}]{2007ASPC..376..127M} {McMullin}, J.~P., {Waters}, B., {Schiebel}, D., {Young}, W., \& {Golap}, K.  2007, in Astronomical Society of the Pacific Conference Series, Vol. 376,
  Astronomical Data Analysis Software and Systems XVI, ed. R.~A. {Shaw},
  F.~{Hill}, \& D.~J. {Bell}, 127

\bibitem[Muxlow et al.(1996)]{1996MNRAS.278..854M} Muxlow, T.~W.~B., Pedlar, A., Holloway, A.~J., Gallimore, J.~F., \& Antonucci, R.~R.~J.\ 1996, \mnras, 278, 854 

\bibitem[Nixon et al.(2012)]{2012MNRAS.422.2547N} Nixon, C.~J., King, A.~R., \& Price, D.~J.\ 2012, \mnras, 422, 2547

\bibitem[Osterbrock \& Shaw(1988)]{1988ApJ...327...89O} Osterbrock, D.~E., \& Shaw, R.~A.\ 1988, \apj, 327, 89 

\bibitem[Ramos Almeida \& Ricci(2017)]{recentReview} Ramos Almeida, C., \& Ricci, C.\ 2017, Nature Astronomy, 1, 679

\bibitem[Sault \& Conway(1999)]{1999ASPC..180..419S} Sault, R.~J., \& Conway, J.~E.\ 1999, Synthesis Imaging in Radio Astronomy II, 180, 419 

\bibitem[Schmitt et al.(2003)]{2003ApJ...597..768S} Schmitt, H.~R., Donley, J.~L., Antonucci, R.~R.~J., et al.\ 2003, \apj, 597, 768 

\bibitem[Tremaine \& Davis(2014)]{2014MNRAS.441.1408T} Tremaine, S., \& Davis, S.~W.\ 2014, \mnras, 441, 140

\bibitem[Tovmassian(2001)]{2001AN....322...87T} Tovmassian, H.~M.\ 2001, Astronomische Nachrichten, 322, 87 

\bibitem[van der Marel \& Franx(1993)]{1993ApJ...407..525V} van der Marel, R.~P., \& Franx, M.\ 1993, \apj, 407, 525 


\end{thebibliography}
\end{document}